\UseRawInputEncoding
\documentclass[12pt,a4paper]{article} 
\pdfoutput=1
\usepackage{jheppub}
\usepackage[T1]{fontenc}
\usepackage{enumerate}
\usepackage{epsfig} 
\usepackage{float} 
\usepackage[caption = false]{subfig}
\usepackage{wasysym} 
\usepackage{mathrsfs} 
\usepackage{amsfonts} 
\usepackage{amsbsy} 
\usepackage{amscd} 
\usepackage{pstricks} 
\usepackage{multirow}
\usepackage{tikz}
\usepackage{color}
\usepackage{slashed,cancel}
\usepackage{array}
\usepackage{tablefootnote}
\usepackage{enumitem}

\usetikzlibrary{arrows,positioning,shapes.geometric} 
\usepackage{color}
\definecolor{myred}{rgb}{0.6,0,0} 
\definecolor{myblue}{rgb}{0,0.2,0.4}
\definecolor{mygreen}{rgb}{0,0.9,0.1}
\definecolor{hc}{rgb}{.9,0.1,0.7}
\definecolor{hcout}{rgb}{.9,0.7,0.9}
\definecolor{Orange}{rgb}{1.,0.65,0.}

\def\MGvATNLO{{\tt {\sc MadGraph5}\_aMC@NLO}}

\title{Probing non-standard $b\bar{b}h$ interaction at the LHC at $\sqrt{s}=13$ TeV} 
\author[a]{Partha~Konar,}
\author[b]{Biswarup~Mukhopadhyaya,}
\author[c]{Rafiqul~Rahaman,}  
\author[b]{and Ritesh~K. Singh}
\affiliation[a]{Physical Research Laboratory,   Ahmedabad - 380009, Gujarat, India}
\affiliation[b]{Department of Physical Sciences, 
	Indian Institute of Science Education and Research Kolkata,  
	Mohanpur, 741246, India}
\affiliation[c]{Regional Centre for Accelerator-based Particle Physics,   Harish-Chandra Research Institute, HBNI, 
Chhatnag Road, Jhusi, Prayagraj 211019, India}
	
\emailAdd{konar@prl.res.in}
\emailAdd{biswarup@iiserkol.ac.in}
\emailAdd{rafiqulrahaman@hri.res.in}
\emailAdd{ritesh.singh@iiserkol.ac.in} 

\abstract{
In the detailed probe of Higgs boson properties at the Large Hadron Collider, and in looking for new physics signatures in the electroweak symmetry breaking sector, the bottom quark Yukawa coupling has a crucial role. We investigate possible departure from the standard model  value of $b\bar{b}h$ coupling, phenomenologically expressed in terms of a modification factor $\alpha_b$,  in $b\bar{b}$-associated production of the $125$-GeV scalar at the high-luminosity LHC. In a next-to-leading order  estimate, we make use of a gradient boosting algorithm to improve in statistical significance upon a cut-based analysis. It is found possible to probe down to $\alpha_b = 3$ with more than $5~\sigma$ significance, with ${\cal L} = 3000$ fb$^{-1}$ and $\sqrt{s}$ = 13 TeV, while the achievable limit at $95\%$ C.L. is $\pm 1.95$.	
}


\begin{document}

\maketitle
\section{Introduction}
\label{sec:into}

Whether the 125 GeV scalar discovered in 2012~\cite{Chatrchyan:2012xdj,Aad:2012tfa} is  `the Higgs' or `a Higgs'
is still an unresolved issue. Most importantly, its interaction strengths
with relatively heavy fermions are not yet known precisely enough, in contrast
to the interaction with gauge boson pairs,  
where the uncertainty is much lesser~\cite{Aad:2019mbh}.
For example, the signal strength  defined as 
$\mu_b = \frac{\sigma (b \bar{b})}{\sigma (b\bar{b})_{SM}}$, where the denominator
corresponds to the rate predicted by the standard model (SM), lies in the range $0.84$ -- $1.24$~\cite{Sirunyan:2018kst}.
Thus there is considerable scope of variation with respect to the standard model value. Here, we propose one way
of reducing this uncertainty, by taking a fresh look at  
$h$-production associated with $b\bar{b}$ at the high-luminosity Large Hadron 
Collider (HL-LHC).

The $b\bar{b}$-associated production of Higgs has been already studied~\cite{Balazs:1998nt,Harlander:2003ai,Dittmaier:2003ej,Dawson:2003kb,Campbell:2004pu,Dawson:2005vi,Wiesemann:2014ioa,Forte:2015hba,Jager:2015hka,Bonvini:2016fgf,Deutschmann:2018avk}, and the
the conclusion is that the rates are too small to make any difference, as far as the SM interaction is concerned. However, the rather large error-bar  
keeps alive the possibility of enhancement  in the presence of new physics.
This is reflected, for example, in two Higgs doublet models (2HDM) where
regions in the parameter space with a large $b$-coupling of the $125$ GeV scalar
are still consistent with all experiments~\cite{Fontes:2015mea}. 
It is important, therefore, to look for clear signatures of such enhancement as the stamp of new physics.

Taking a model-independent standpoint, let us parametrize 
the modification factor for the $b\bar{b}h$ interaction strength by
$\alpha_b$, treated here as real, as 
\begin{equation}
\alpha_b = \frac{y_b}{(y_b)_{\text{SM}}}.
\end{equation}
Here $(y_b)_{\text{SM}}=\sqrt{2}m_b/v$ is the SM bottom Yukawa coupling  ($v=246$ GeV
is the vacuum expectation value), and $y_b$ is the bottom Yukawa coupling in a new physics model.
The analysis of Higgs-$p_T$ data with $\int {\cal{L}}dt = 35.9 fb^{-1}$~\cite{Sirunyan:2018sgc} 
already restricts $\alpha_b$  and $\alpha_c$ (its analogue for the charm)
as $-1.1 \le  \alpha_b \le +1.1, ~-4.9 \le \alpha_c \le +4.8$ at $95\%$ C.L.. However, no
other non-standard  Higgs interaction is allowed in such an analysis, and thus
the contributions to $h \rightarrow Z Z, \gamma \gamma$ bring in stringent 
constraints. However, in the absence of this restrictive assumption and
allowance for `nuisance parameters' relaxes the corresponding ranges to
$\left[-8.5,18.0\right]$ for $\alpha_b$ and 
$\left[-33.0,38.0\right]$ for $\alpha_c$. A more recent study~\cite{Cepeda:2019klc} in the context of
the high-luminosity LHC, running upto 3000 $fb^{-1}$, yields the projected
limits as $-2.0 \lesssim  \alpha_b \le +4.0,~ -10.0 \le \alpha_c \le +10.0$ at $95\%$ C.L.,
once other non-standard interactions are not forbidden, 
and the branching ratios in
the $ZZ$ and $\gamma\gamma$ channels are not used as prior constraints.

We show here that $\alpha_b$ can be pinned down to an even shorter range
by considering $pp \rightarrow b \bar{b} h$ in the high-luminosity run.
In this channel, significant enhancement takes place 
at the production level itself for large $\alpha_b$. This is of
advantage, since the level of enhancement 
does not saturate with increasing $\alpha_b$, unlike the effect 
on the branching ratio in the $b\bar{b}$ channel when the anomalous
$b\bar{b}$ shows up in decays alone.

The resulting signal, where one looks for four $b$-jets
with two of them close to the $h$-peak, is jacked up substantially
for $\alpha_b \rightarrow 3.0$. However, it is also plagued by backgrounds,
including four $b$-jets from QCD, $b\bar{b}Z$ production, and also QCD
production of $2b2c$, with two $c$-quark jets faking $b$'s. 

The backgrounds pose larger next-to-leading order (NLO) QCD corrections strengths than that of the  signal, thus
making the signal significance smaller at NLO   than the leading-order (LO) values.
Our analysis
reveals how the resulting loss in signal significance due to the NLO QCD effects can be 
ameliorated by adopting an algorithm based on Boosted Decision Trees (BDT)---
in particular, the gradient boosting technique. 

In section~\ref{sec:signa-backgrounds}, we provide an outline of the framework to operate within,
with $\alpha_b$ (and $\alpha_c$) taken as purely phenomenological parameters,
with no bar {\em prima facie} on other non-standard interactions. We also discuss the signal and all major irreducible background processes involved in the present analysis.
Sections~\ref{sec:cut} and~\ref{sec:xgboost} contains, respectively, report  on  cut-based and BDT-based machine learning 
analyses. We summarise and conclude in section~\ref{sec:summary}.

\section{The parametrisation of anomalous couplings}\label{sec:signa-backgrounds} %

\begin{figure}[tbh]
	\centering
	\includegraphics[width=0.8\textwidth]{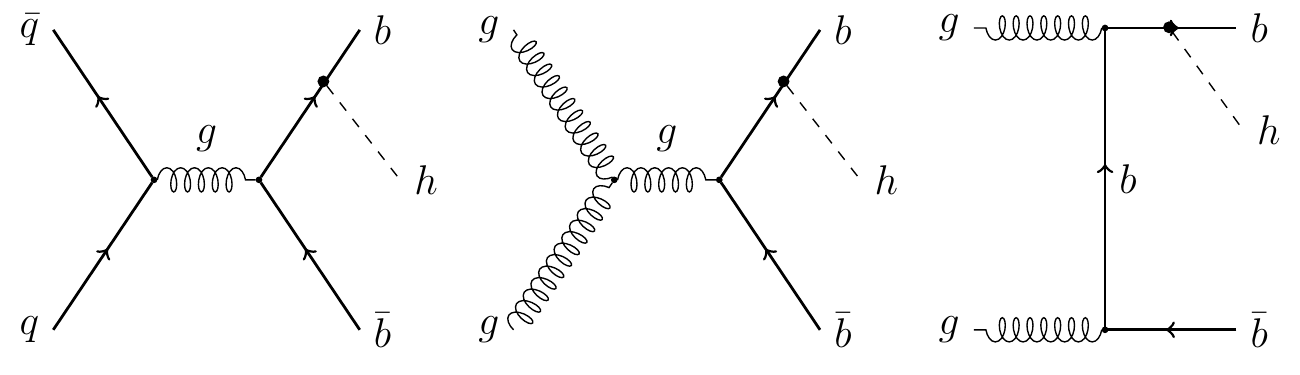}
	\caption{\label{fig:feynman-bbh} Representative Feynman diagram for the $b\bar{b}h$ production at the LHC at leading order (LO). The non-standard $b\bar{b}h$ coupling is shown by the blobs.  Higgs decays further through $h\to b\bar{b}$. The second blob in such vertex are not shown in the diagrams.}
\end{figure}

\begin{figure}[bth!]
	\centering
	\includegraphics[width=0.6\textwidth]{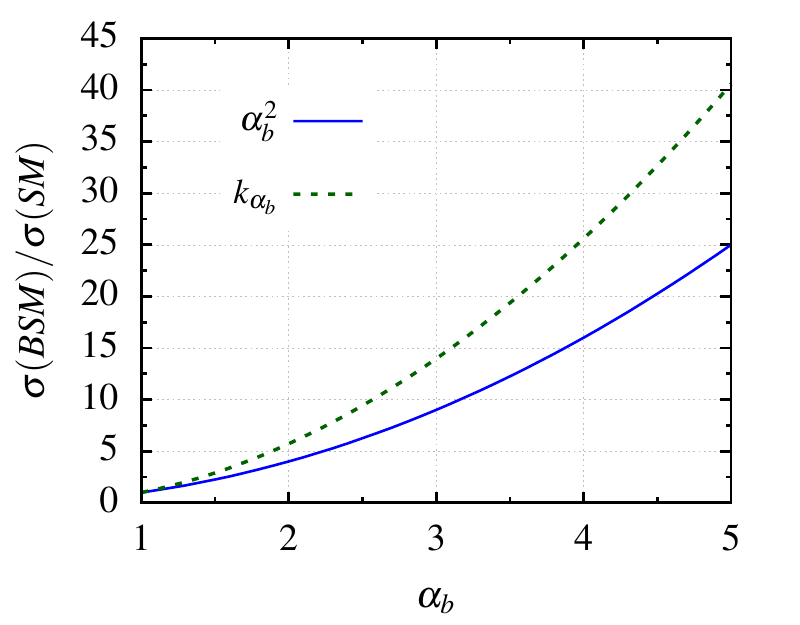}
	\caption{\label{fig:bbh4b-enhancment}  The enhancement factor received in the signal cross section over SM, as defined at Eq. \ref{eq:enhancement} is shown for variation of modification factor in $ b\bar{b}h$ interaction strength $\alpha_b$. }
\end{figure}
We are interested in the $b\bar{b}$-associated Higgs  production followed by the Higgs decaying to a pair of $b$ at the LHC, thus resulting in a $4b$ final state.  The representative Feynman diagrams for the $b\bar{b}h$ production at the LHC are shown in Figure~\ref{fig:feynman-bbh}. The $h\to b\bar{b}$ decay is not shown in the diagrams. 
The $\alpha_b$ also appears in the decay vertex of the $h\to b\bar{b}$ apart from the production process. The total cross section of the signal with $\alpha_b$ will be,
\begin{equation}
\sigma_{b\bar{b}h\to 4b}(\alpha_b) = \alpha_b^2 \, \sigma_{b\bar{b}h}(SM)\times  \dfrac{\alpha_b^2 \, \Gamma(h\to b\bar{b})}{\Gamma_h(\alpha_b)}
\end{equation}
enhancing the SM cross section by a factor of 
\begin{equation}\label{eq:enhancement}
k_{\alpha_b}=\dfrac{\sigma_{b\bar{b}h\to 4b}(\alpha_b) }{\sigma_{b\bar{b}h\to 4b}(SM) } = \alpha_b^2 \, \dfrac{\alpha_b^2\Gamma_h(SM)}{\Gamma_h(\alpha_b)}.
\end{equation} 

\begin{figure}[tbh]
	\centering
	\includegraphics[width=0.6\textwidth]{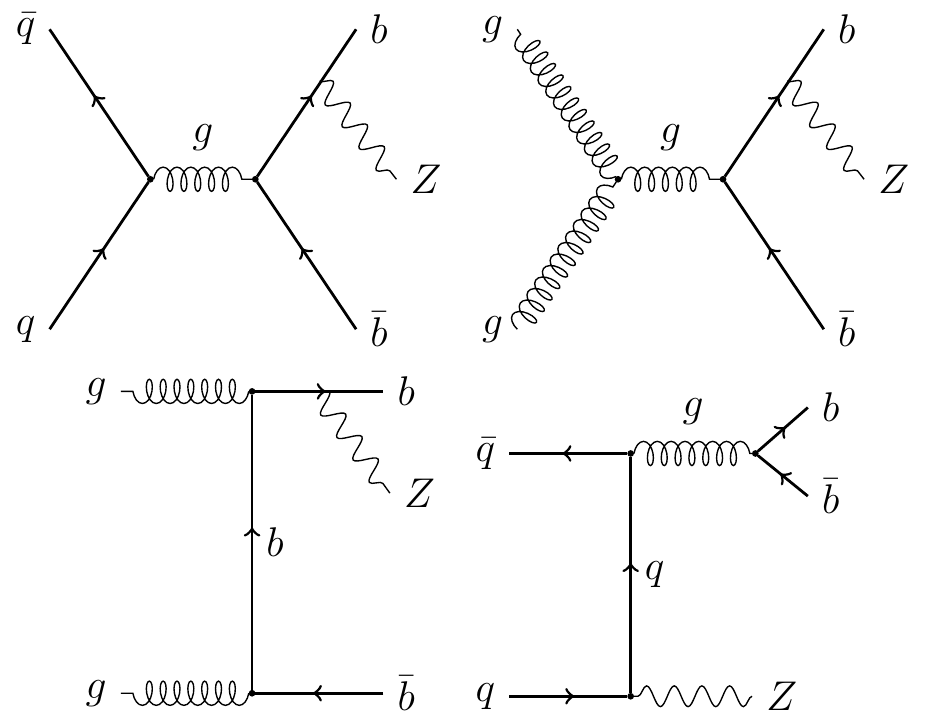}
	\caption{\label{fig:feynman-bbZ} Representative Feynman diagram for the topologically very similar background process through $b\bar{b}Z$ production at the LHC at LO. }
\end{figure}
The enhancement factor for the signal cross section over the SM is shown in Figure~\ref{fig:bbh4b-enhancment} with varying $\alpha_b$ for $B(h\to b\bar{b}) \approx60\%$~\cite{Tanabashi:2018oca}. The solid/{\em blue} line represents the factor when the new physics effect is accounted for only in the production part; the dashed/{\em green} line, however, represents the enhancement factor when the new physics is accounted for in both production as well as in the decay process.

\begin{table}[tbh!]
	\centering
	\renewcommand{\arraystretch}{1.50}
	\begin{tabular}{|c|c|} \hline
		Process &LO  cross section(pb)  \\ \hline
		$bbh\to 4b$  ($\alpha_b=1$) &  $5.386_{-21.6\%}^{+29.5\%}(\text{scale})\pm 12.4\%(PDF)\times 10^{-3}$ \\ \hline
		$bbZ \to 4b $& $2.082_{-20.1\%}^{+26.9\%}(\text{scale})\pm 11.5\%(\text{PDF})$ \\ \hline
		QCD-$4b $ & $118.194_{-35.5\%}^{+60.9\%}(\text{scale})\pm 12.4\%(\text{PDF})$ \\ \hline
		QCD-$2b2c$ & $636.098_{-35.4\%}^{+60.6\%}(\text{scale})\pm 12.6\%(\text{PDF})$ \\ \hline
		$hZ \to 4b $& $0.01764_{-5.8\%}^{+4.7\%}(\text{scale})\pm 6.1\%(\text{PDF})$ \\ \hline
	\end{tabular}
		\caption{\label{tab:lo-cross-sections}Parton level  cross sections of signal and the major background processes at the leading order (LO), after applying the generator level event selection cuts of
	$p_T(b)>30$ GeV, $\Delta R (b,b) > 0.4$, $\eta_b < 2.5$ using the scale choice of $\mu_R=\mu_F=\mu_0=(2 m_b + m_h)/2$.
	}
\end{table}

The major backgrounds to the $4b$ final state comes  from QCD $4b$-jets, QCD $2b2c$  with the $c$-quarks faking as $b$-jets, $b\bar{b}Z$ production, and $hZ$ production. We ignore the QCD $4j$ process as the  probability of four light jets faking as four $b$-jets is insignificantly small.
The background $b\bar{b}Z$, Feynman diagrams shown in Figure~\ref{fig:feynman-bbZ}, has the same topology as the signal $b\bar{b}h$, and thus expected to be irreducible from the signal.  The QCD  and the $hZ$ backgrounds, however, are expected to be reducible for having a different topology than that of the signal.  The leading order  cross sections for the signal  and the backgrounds for the $4b$ final state estimated in the \MGvATNLO~v2.6.4 (mg5\_aMC)~\cite{Alwall:2014hca} package with a generator level cuts of $p_T(b)>30$ GeV, $\Delta R (b,b) > 0.4$, and $\eta_b < 2.5$ are presented in the Table~\ref{tab:lo-cross-sections}. We use a fixed renormalization ($\mu_R$) and factorization ($\mu_F$) 
scale of $\mu_R=\mu_F=\mu_0=(2 m_b + m_h)/2$ for the signal as well as for the backgrounds motivated by the $b\bar{b}h$ production topology.  The scale uncertainties, shown in Table~\ref{tab:lo-cross-sections},  are estimated
by varying the $\mu_R$ and $\mu_F$ in the range of $0.5\mu_0 \le \mu_R , \mu_F \le 2\mu_0$, with the constraint $0.5\le\mu_R/\mu_F\le2$.
We use the NNPDF3.0~\cite{Ball:2014uwa} sets of parton distribution functions (PDFs) with $\alpha_s(m_Z)=0.118$ for our calculations.
A branching ratio of $60\%$ is used for the $h\to b\bar{b}$ decay~\cite{Tanabashi:2018oca} with $m_h=125$ GeV. 

\section{Cut-based analysis}\label{sec:cut}
We generated events for  the signal and the backgrounds in  mg5\_aMC   at LO and NLO with chosen renormalization and factorization scale. The QCD-$2b2c$ background, however, is generated only at LO, and it is used for NLO analysis with a $k$-factor of $1.4$ taken from the QCD-$4b$ background.   The showering and hadronization of the events are performed by {\tt PYTHIA8}~\cite{Sjostrand:2014zea} followed by the detector simulations by {\tt Delphes}-3.4.2~\cite{deFavereau:2013fsa}. We estimated the expected number of events with four $b$-tagged jets for the signal with $\alpha_b=3$  and the backgrounds after detector simulations at an integrated luminosity of ${\cal L} = 3000$ fb$^{-1}$   for the following two kinematical regions:
\begin{eqnarray}\label{eq:cuts-region}
&\text{Event selection (\textsf{cut1})}&:~~  p_T(b)>20~\text{GeV},~\Delta R (b,b) > 0.5,~\eta_b < 2.5,\\
&\text{Event selection (\textsf{cut2})}&:~~  \textsf{cut1} + \text{ at least one }m_{bb}\in[100,150]~\text{GeV}
\end{eqnarray}
and present them in Table~\ref{tab:cut-based-significance} for $\mu_R=\mu_F=\mu_0=(2 m_b + m_h)/2$. 
\begin{table}[tb]
	\centering
	\renewcommand{\arraystretch}{1.50}
	\begin{tabular}{|c|c|c|c|c|c|} \hline
		\multicolumn{2}{|c|}{Signal \& background}&\multicolumn{2}{c|}{ No. of events @ LO}&\multicolumn{2}{c|}{ No. of events @ NLO}\\ \cline{3-6} 
		\multicolumn{2}{|c|}{process} & \textsf{cut1} & 	 \textsf{cut2}& \textsf{cut1} & 	 \textsf{cut2}\\ \hline \hline 
		$S    $ : & $bbh\to 4b$  & 		$ 33511$ & $ 30867$ & $38895 $& $34946.8 $    \\ \hline 
		$B_1$ : & $bbZ \to 4b $ & $846715$ & $682871$& $1.67229\times 10^6 $& $1.33163\times 10^6 $\\ \hline 
		$B_2$ : & QCD - $4b $ & $ 4.24088\times 10^7 $ & $3.36035\times 10^7$& $6.81198\times 10^7 $& $5.07642\times 10^7 $\\ \hline 
		$B_3$ : & QCD-$2b2c$  & $1.53389\times 10^7$ &$1.1986\times 10^7$& $2.15198\times 10^7 $& $1.68568\times 10^7 $\\ \hline 
		$B_4$ : & $hZ \to 4b $ & $7817$ &$7168$& $20177 $& $18244 $\\ \hline \hline
		\multicolumn{2}{|c|}{Significance ($\frac{S}{\sqrt{B}}$) }
		 & $4.38$ &$4.54$& $4.07 $& $4.21 $\\ \hline 
	\end{tabular}
\caption{\label{tab:cut-based-significance} 
Expected number of events for the signal as well as the backgrounds and signal significance ($S/\sqrt{B}$) 
 at  ${\cal L} = 3000$ fb$^{-1}$ with $\alpha_b=3$ at LO as well as at NLO for $\mu_R=\mu_F=\mu_0=(2 m_b + m_h)/2$ for two cuts region given in Eq.~\ref{eq:cuts-region}.}
\end{table}
For the \textsf{cut2}, we select events with at least one  $b$ pair with invariant mass in the range  
$[100,150]$, thus emulating a Higgs candidate. We calculate the signal significance, defined by $S/\sqrt{B}$ with $S$ being signal events and $B$ being total background events, for the two cut region, and they are shown in the lowest row of Table~\ref{tab:cut-based-significance}. 
A $b\bar{b}h$ signal with $\alpha_b=3$ can be observed with a significance of $4.54$ ($4.21$) at LO (NLO) in the \textsf{cut2} region at an integrated luminosity of ${\cal L} = 3000$ fb$^{-1}$ for renormalization and factorization scale of $\mu_R=\mu_F=\mu_0=(2 m_b + m_h)/2$. The signal significance for other renormalization and factorization scales namely $\mu_R=\mu_F=\mu_0/2,~2\mu_0$  are also shown in the next section for comparison.
The QCD corrections for the signal being  much smaller compared to the same for the QCD backgrounds, and the shape of the distributions of the variables being similar for LO and NLO, the signal significance is smaller at NLO compared to the  LO result.
Other than the \textsf{cut2} regions, cuts such as $p_T$, $H_T$, $m_{4b}$, $\cancel{E}_T$  on the $b$-jets do not improve the signal significance. 
These variables, however, in certain combinations, may improve the signal significance, which we explore with the gradient boosting technique in the next section.

\section{Analysis based on the gradient boosting technique}\label{sec:xgboost}
After estimating a  maximally achievable signal significance with a simple cut-based analysis (\textsf{cut2} in Eq.~(\ref{eq:cuts-region})), we further explore the 
possibility of improving the significance by a Machine Learning technique namely {\em Gradient Boosted Decision Trees} (gradient BDT)~\cite{Chen:2016btl} by employing various kinematical variables. We use the package {\tt XGBoost}~\cite{Chen:2016btl} as a toolkit for the gradient boosting.  We construct these following kinematical features as input for the gradient boosting:
\begin{itemize}
	\item Transverse momentum of each of the four leading b-tagged jets $p_T(b_i)$ (4 variables),
	\item Total invariant mass of all four leading b-tagged jets $m_{4b}$ and inclusive variables, such as, missing transverse momentum $MET$, global mass scale variable $H_T$ (3 variables),
	\item Set of all b-jet pair invariant masses $m_{b_i b_j}$, and b-pair transverse momentum $p_T(b_i b_j)$  from all four leading b-tagged jets (12 variables),
	\item $\theta$-angle (measured w.r.t. the boost of $4b$-system) and pseudo-rapidity of each b-tagged jet $\cos\theta(b_i)$, $\eta_{b_i}$  (8 variables),
	\item Angular and azimuthal angle separation between set of all b-jet pairs $\Delta R(b_ib_j)$, $\Delta\phi(b_ib_j)$ (12 variables),
	\item Angular and azimuthal angle separation between the `invariant-mass based reconstructed' Higgs candidate (composed of two b-tagged jets - so called $b_1$ and $b_2$) from other two b-tagged jet candidates $\Delta R(hb_3)$, $\Delta R(hb_4)$, $\Delta\phi(hb_3)$, $\Delta\phi(hb_4)$ (4 variables),
	\item Angular and azimuthal angle separation between the `angular-separation based reconstructed' Higgs candidate (composed of two b-tagged jets - so called $b_1$ and $b_2$) from other two b-tagged jet candidates $\Delta R(h^\prime b_3^\prime )$, $\Delta R(h^\prime  b_4^\prime )$ , $\Delta R(b_3^\prime  b_4^\prime )$, $\Delta\phi(h^\prime b_3^\prime )$, $\Delta\phi(h^\prime b_4^\prime )$,   $\Delta\phi(b_3^\prime b_4^\prime )$ (6 variables).
\end{itemize}
The number in parentheses at the end of each item represents the total number of features in each item, giving a total of $49$ features. The  features are reconstructed as follows: The Higgs candidate ($h$) is reconstructed with the $b$-pair close to $125$ GeV invariant mass. These two $b$'s are labelled as $b_1$ and $b_2$ ordered by their $p_T$. The other two $b$'s are labelled as $b_3$ and $b_4$, also ordered by their $p_T$. On the other hand, the primed Higgs candidate ($h^\prime$) is reconstructed using the lowest $\Delta R$ of the $b$-pairs. The $b_i^\prime$ are labelled in a similar way as done in the un-primed case. 

\begin{figure}[tb!]
	\centering
	\includegraphics[width=0.48\textwidth]{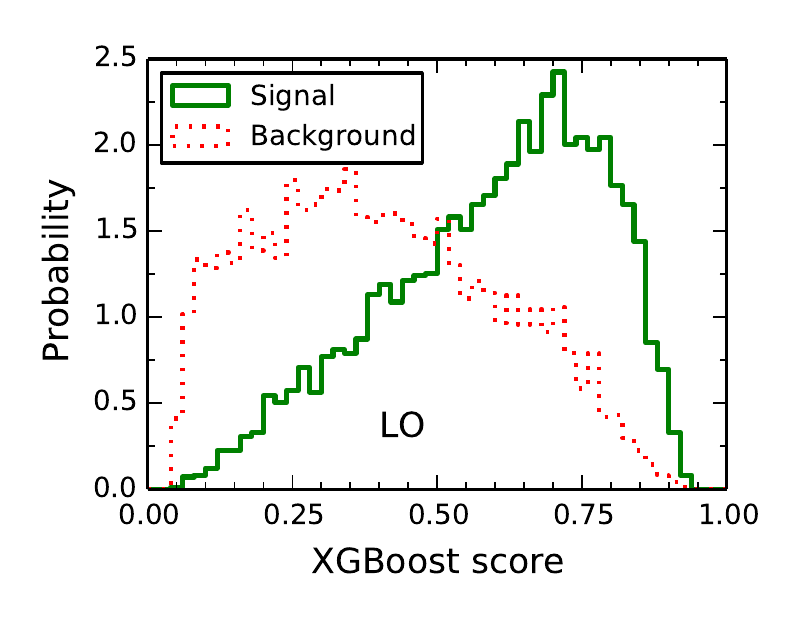}
	\includegraphics[width=0.48\textwidth]{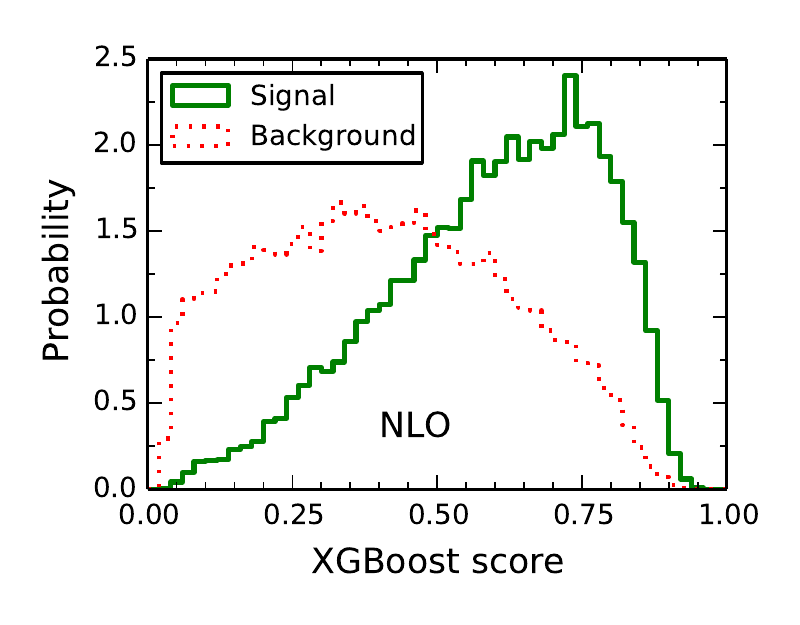}
	\includegraphics[width=0.48\textwidth]{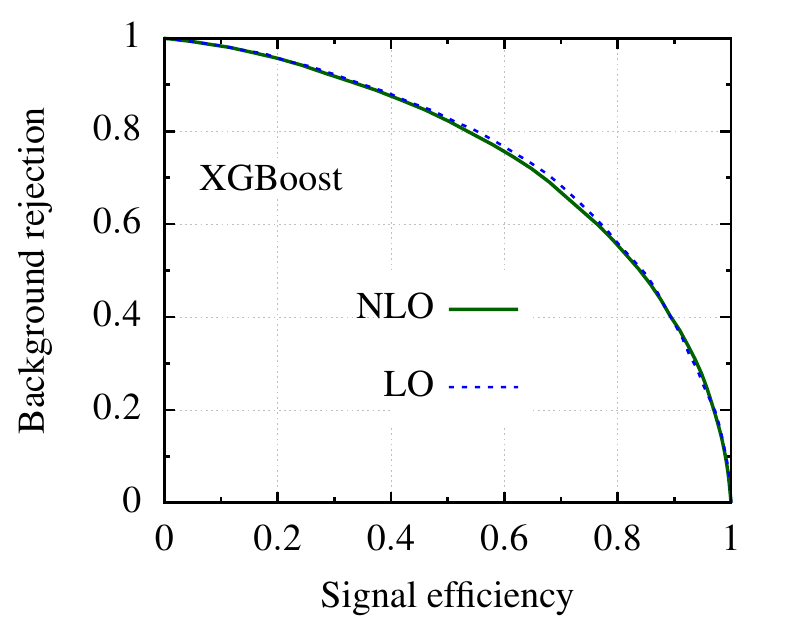}
	\includegraphics[width=0.48\textwidth]{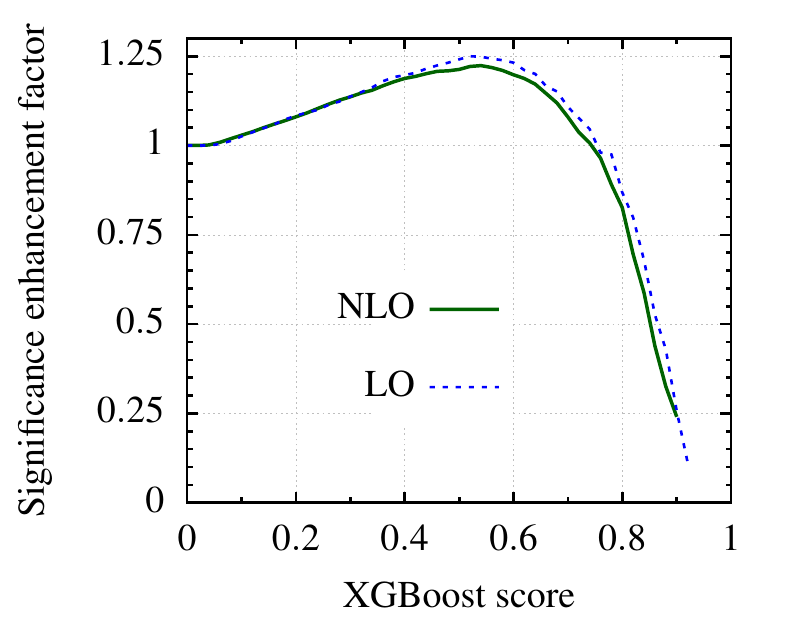}
	\caption{\label{fig:Ml-result} 
	Normalised distribution of BDT response for the signal and background processes representing relative separability based on the XGBoost score cut for LO ({\em top-left}) and NLO ({\em top-right}). Corresponding performance in terms of ROC (receiver operating characteristic) curve shown in {\em bottom-left} curve. Also the {\em right-bottom} plot is for significance enhancement factor depending on the XGBoost score cut for $\mu_R=\mu_F=\mu_0=(2 m_b + m_h)/2$.
}
\end{figure}

We use an equal number of events for the signal and background events to classify them using the {\em train} module of {\tt XGBoost}. The backgrounds are mixed with the  ratio of their corresponding rates  in \textsf{cut2} region given  in Table~\ref{tab:cut-based-significance}.  We use $80\%$ of the total dataset for training purposes and the rest $20\%$ for testing purposes.
At first, we vary the XGBoost parameters 
to obtain a combination of them for a maximum accuracy to classify the signal and the backgrounds. We obtain a maximum accuracy of $69.13\%\pm0.44 \%~(1\sigma)$ for LO events and $68.06\%\pm0.23 \%~(1\sigma)$ for NLO events at $\mu_R=\mu_F=\mu_0$ for the following combination of the parameters' values \cite{xgboost_param}: 
\begin{itemize}
\item \textsf{Step size shrinkage:} $\eta             = 0.1$,
\item \textsf{Maximum depth of a tree:} \text{{\em max\_depth}}       = 50,
\item \textsf{Subsample ratio of the training instances:} \text{{\em subsample }}      = 0.9,
\item \textsf{subsample ratio of columns when constructing each tree:} \text{{\em colsample\_bytree}}= 0.3,
\item \textsf{Minimum loss reduction required to make a further partition on a leaf node of the tree:} $\gamma= 1.0$,
\item \textsf{L2 regularization term on weights:} $\lambda=50.0$,
\item \textsf{L1 regularization term on weights:} $\alpha=1.0$,
\item \textsf{Number of parallel trees constructed during each iteration:} \text{{\em num\_parallel\_tree}}=8.
\end{itemize}

\begin{table}[tb!]
	\centering
	\renewcommand{\arraystretch}{1.50}
	\begin{tabular}{|c|c|c|c|c|c|} \hline
	\multicolumn{2}{|c|}{} &{\textbf{@LO}} &\multicolumn{3}{c|}{\textbf{@NLO} }\\ \hline
	\multicolumn{2}{|c|}{Scale choice ($\mu_R=\mu_F=\mu_0)$}	          &  $\mu_0$ & $\mu_0/2$ & $\mu_0$ &$2\mu_0$ \\ \hline\hline 
	\multicolumn{2}{|c|}{Cut-based significance ($\frac{S}{\sqrt{B}}$)}       &	4.54	     &$2.92$       &  $4.21 $.  & $5.06$ \\ \hline\hline 
	\multirow{4}{*}{\rotatebox{90}{XGBoost}}
	&	Signal efficiency  ($\epsilon_S$)  &    67.7\%            &$74.3\%$  &$70.8\%$ & $67.3\%$ \\ \cline{2-6} 
	&	Brackground rejection  ($\bar{\epsilon}_B$) &    70.7\%     &$63.1\%$ &$65.9\%$& $68.5\%$ \\ \cline{2-6} 
	&	Enhancement factor ($\frac{\epsilon_S}{\sqrt{1-\bar{\epsilon}_B}}$) &   1.25  &$1.22$ &$1.2$& $1.2 $\\ \cline{2-6}  
	&	Maximum significance ($\frac{S}{\sqrt{B}}\times \frac{\epsilon_S}{\sqrt{1-\bar{\epsilon}_B}}$) &  5.67  &$3.56$  & $5.05$ & $6.07 $ \\ \hline 
	\end{tabular}
\caption{ \label{tab:cut-based-xgboost}  \label{tab:muR-muF-variation}
Comparison of signal efficiency $\epsilon_S$, background rejection $\bar{\epsilon}_B \equiv (1 - \epsilon_B)$, efficiency factor and significance between the leading order (LO) calculation and the next to leading order (NLO). Both cut and count based and XGBoost analysis results are shown at the luminosity  ${\cal L} = 3000$ fb$^{-1}$  after the final event selection as in \textsf{cut2}, for a moderate value of modification factor, 
$\alpha_b=3$.  Also the effect of variation in renormalization and factorization scale choice at the NLO pointed out in additional columns. Note for the XGBoost analysis, results are shown for a choice of XGBoost score cut where the maximum significance is achivable.
%
}
\end{table}
 The final probability distributions of the  output of the BDT network (XGBoost score) for the signal and the total backgrounds are shown in Figure~\ref{fig:Ml-result} in the {\em top-row} to show their separability for LO ({\em left}) as well as for NLO ({\em right}). The signal efficiency versus the background rejection curves for LO and NLO are shown in Figure~\ref{fig:Ml-result} in the {\em left-bottom-panel}. The larger the area under the curves of signal efficiency versus the background rejection, the better is the separability between the signal and the background.  Compared to the LO, the NLO events are less distinguishable, thus reduces the background rejection.
 The XGBoost score cut is varied to obtain the signal efficiency $\epsilon_S$, background rejection $\bar{\epsilon}_B \equiv (1 - \epsilon_B)$
 for the maximum factor by which the significance can be improved. The significance enhancement factor ($\frac{\epsilon_S}{\sqrt{1-\bar{\epsilon}_B}}$) w.r.t. the XGBoost score cut is shown in the {\em right-bottom-panel} in Figure~\ref{fig:Ml-result}.
 For a XGBoost score of $0.52$ ($0.50$), $67.7\%$ ($70.8\%$) signal remains  rejecting a total of $70.7\%$ ($65.9\%$) background  for LO (NLO) events, thus maximally enhancing the signal significance by a factor of $1.25$ ($1.2$). 
 Thus the total significance after the BDT analysis becomes $4.54\times 1.25 =5.67$ ($4.21\times 1.2=5.058$) at LO (NLO). The combined result of cut-based and XGBoost at NLO are shown in  Table~\ref{tab:cut-based-xgboost} in the second and fourth column for LO and NLO, respectively. Table~\ref{tab:cut-based-xgboost}  also contains result for renormalization and factorization scale choice $\mu_0/2$ and $2\mu_0$ for NLO; The reason being discussed below.
 
 


 \begin{figure}[tb!]
	\centering
	\includegraphics[width=0.49\textwidth]{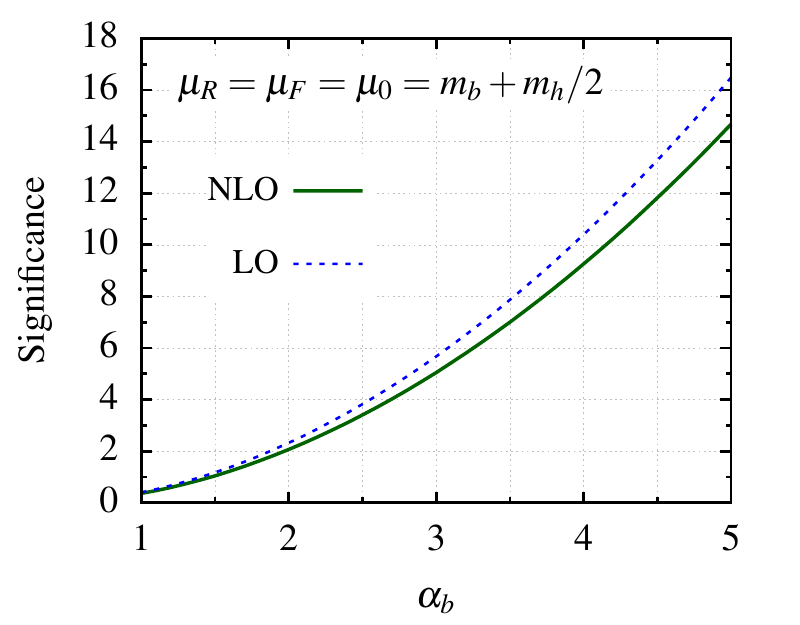}
	\includegraphics[width=0.49\textwidth]{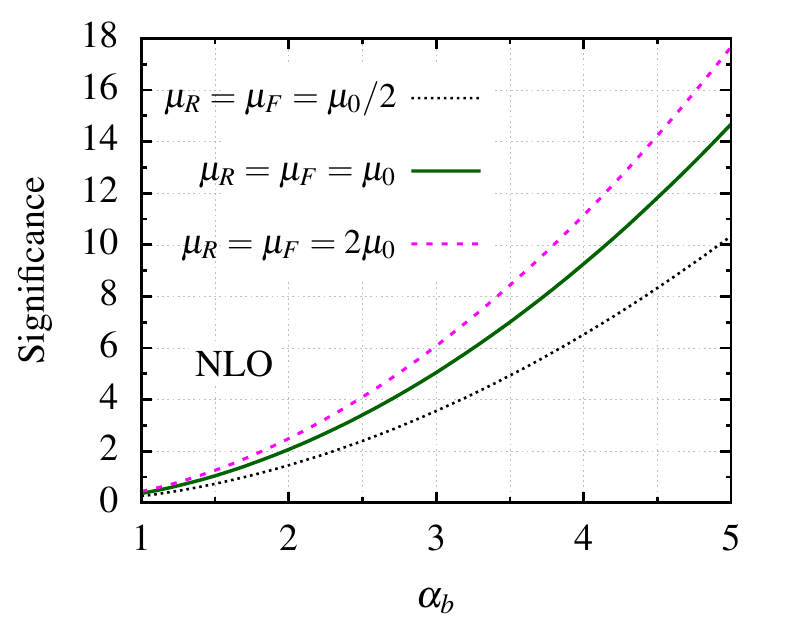}
	\caption{\label{fig:alphab-vs-significance}  Achievable total significance as a function of modification factor for the $hb\bar{b}$ interaction strength $\alpha_b$ at LO and NLO in {\em left-panel} with renormalization and factorization scales at $\mu_0=m_b+m_h/2$ for  ${\cal L} = 3000$ fb$^{-1}$. 
The {\em right-panel} shows the scale variance of such significance at NLO considering two extreme cases of changing both the scales with a factor of half and a factor of two. }
\end{figure}


The QCD correction strengths and the shape of the distributions for the kinematical variable change as the renormalization and factorization scale are changed for the signal as well as for backgrounds.  As a result, our results in cut-based as well as in BDT-based analysis are expected to be different for different $\mu_R$ and $\mu_F$. So, we repeat all analyses for two extreme cases of $\mu_R$ and $\mu_F$ with a factor of half and one, i.e.,  $\mu_R=\mu_F=\mu_0/2,~\mu_0,~2\mu_0$ apart from $\mu_0$ and obtain the results. The results are shown in Table~\ref{tab:muR-muF-variation} for $\mu_R=\mu_F=\mu_0/2,~\mu_0,~2\mu_0$. The results for $\mu_0$ are repeated for comparison.  
The QCD correction strength increase for the signal as the scale choices are doubled to $2\mu_0$, while it decreases as the  scale choices are reduced to $\mu_0/2$.  However, the QCD corrections remain roughly the same for the backgrounds, specially for the dominant $4b$ QCD background. As a result, 
as the scale choices are doubled, the signal significance improves by  $25\%$, but it decreases when the scale choices are halved  at the cut-based analysis.
 In the XGBoost result, the significance enhancement factor, however, increase a little due to the increase in signal efficiency for lower-scale choices. 

Till now we have shown the results for $\alpha_b=3$, i.e., for a fixed value of the new physics parameter.  The total signal significance, including cut-based and XGBoost, are  computed for varying $\alpha_b$, and they are shown in Figure~\ref{fig:alphab-vs-significance} for $\mu_R=\mu_F= \mu_0$ at LO and NLO in {\em left-panel} at  ${\cal L} = 3000$ fb$^{-1}$. The {\em right-panel}  in Figure~\ref{fig:alphab-vs-significance} shows the comparison of signal significance for three different scale choices namely $\mu_0/2$, $\mu_0$, and $2\mu_0$ at NLO for the same luminosity ${\cal L} = 3000$ fb$^{-1}$.
The limits on  $\alpha_b$ is obtained to be $\pm 1.95$ at $95\%$ C.L. at NLO for $\mu_R=\mu_F=\mu_0$, see Figure~\ref{fig:alphab-vs-significance} {\em left-panel}.
The limits on $\alpha_b$ is tighter for higher scale choices and weaker for lower scale choices, as can be seen in the {\em right-panel}.

It appears that the strengths of QCD corrections for the backgrounds are always higher than that for the signal for a range of renormalization and factorization   scales, thus making the signal significance smaller for NLO than LO for both cut-based and XGBoost analysis. 
A $5\sigma$ discovery significance is achievable for a moderate value of $\alpha_b=3$ at a projected luminosity of $3000$ fb$^{-1}$ at the LHC.

\section{Summary and conclusions}
\label{sec:summary}
While LHC is emphatically looking for any indication of elusive new physics, hints of that can already be hidden in our Higgs data. Precision measurements of Higgs coupling with third-generation quarks are thus crucial in indirect probes on the physics beyond the standard model. In this present work, we probe the non-standard $hb\bar{b}$ coupling parametrized in a model-independent standpoint in $b\bar{b}$-associated production of Higgs. We point out the importance and effectiveness of this channel in uncovering the modification factor in $hb\bar{b}$ interaction strength.


With a detailed detector level simulation, we devised a phase space region to emulate a Higgs peak in the signal and also to regulate the background processes. 
We obtain a moderate signal significance showing the outcome both at LO as well as at NLO for a choice of modification factor $\alpha_b=3$ at high luminosity  
LHC. This cut-based significance is further refined upon by gradient boosted techniques later on. Overall, the NLO result is slightly weaker than that of LO.
We also investigate the effect of (renormalization and factorization) scale variation on the results at NLO and observe a significant variation, with better results at relatively higher scale values.
The limit on $\alpha_b$, which is $\pm 1.95$ at $95\%$ C.L. for ${\cal L} = 3000$ fb$^{-1}$, is surpassing the existing results in literature.

During the concluding stage of study, we came across reference~\cite{Grojean:2020ech}, where an anomalous $hb\bar{b}$ interaction has been investigated via the Higgs decay channel $h\to \gamma\gamma$, in the context of higher energies and luminosities than those envisioned currently for the HL-LHC. Our study differs from theirs in several ways. First, by concentrating on the $b\bar{b}$ decay mode, one expects  considerably larger event rates. Secondly, it also entails more severe backgrounds. 
Thirdly, next-to-leading order QCD effects are more non-trivial, not only for the backgrounds but for the signal as well. We have shown how to overcome the 
second and third issues, especially with the help of gradient boosting techniques, and thus improve upon hitherto estimated levels of constraining 
$\alpha_b$ at the HL-LHC, with $\int {\cal L}dt = 3000$ fb$^{-1}$.

\section{Acknowledgement}
\label{sec:ack}
The work of PK is supported by Physical Research Laboratory (PRL), Department of Space, Government of India. 
The work of RR is partially supported by funding available from the Department of Atomic Energy, Government of India, for the Regional
Centre for Accelerator-based Particle Physics (RECAPP), Harish-Chandra Research Institute.
 The work of RKS is partially supported by SERB, DST, Government of India through the project EMR/2017/002778.  
 
\providecommand{\href}[2]{#2}
\addcontentsline{toc}{section}{References}
\bibliography{References}
\bibliographystyle{utphysM}

\end{document}